\documentclass{PoS}
\usepackage{amsmath,amsfonts,amssymb}

\newcommand{\gm}{\gamma^\mu}
\newcommand{\smn}{\sigma^{\mu \nu}}
\newcommand{\Wmn}{W_{\mu \nu}}
\newcommand{\vl}{V_L}
\newcommand{\vr}{V_R}
\newcommand{\gl}{g_L}
\newcommand{\gr}{g_R}
\newcommand{\DMD}{\overleftrightarrow{D^\mu}}

\title{Effective operators in top physics}

\ShortTitle{Effective operators in top physics}

\author{\speaker{Juan Antonio Aguilar-Saavedra}%
         \thanks{This work has been partially supported by a MICINN Ram\'on y Cajal contract and by projects FPA2006-05294 (MICINN), FQM 101 and FQM 437 (Junta de Andaluc\'{\i}a), CERN/FP/83588/2008 (FCT), and MRTN-CT-2006-035505 (European Union).}\\
        University of Granada \\
        E-mail: \email{jaas@ugr.es}}

\abstract{Top interactions beyond the Standard Model can be conveniently described in the framework of gauge-invariant effective operators. We briefly review the general form of the fermion-fermion-gauge boson and fermion-fermion-Higgs interactions arising from dimension-6 operators, focusing on the top quark. We discuss in detail an expansion in powers of $1/\Lambda^2$, where $\Lambda$ is the new physics scale, to consistently calculate observables in terms of top anomalous couplings.}

\FullConference{35th International Conference of High Energy Physics\\
                 July 22-28, 2010\\
                 Paris, France}

\begin{document}

\section{Introduction}
One of the main motivations for detailed studies of the top quark properties is to search for indirect effects of new physics not directly observable. Heavy physics above the electroweak scale can be parameterised using an effective Lagrangian
\begin{equation}
\mathcal{L}^\mathrm{eff} = \sum \frac{C_x}{\Lambda^2} O_x + \dots \,,
\label{ec:effL}
\end{equation}
where $O_x$ are dimension-6 operators invariant under the Standard Model (SM) gauge symmetry, $\Lambda$ the new physics scale and the dots stand for dimension-8 and higher-order operators, usually ignored. Dimension-6 operators were classified in \cite{Buchmuller:1985jz} and an operator basis was given, which can be used to parameterise any new physics contribution up to higher orders. Later, it has been found~\cite{Grzadkowski:2003tf,AguilarSaavedra:2008zc} that some of these operators are actually redundant and can be dropped from the list.
This simplification is very useful for phenomenology, because it implies that (i) the most general fermion-fermion-gauge boson interactions arising from dimension-6 operators contain at most $\gamma^\mu$ and $\sigma^{\mu \nu} q_\nu$ terms, where $q$ is the boson momentum; (ii) the most general fermion-fermion-Higgs interactions only contain scalar and pseudo-scalar terms. In the particular case of the top quark we have, for example, the most general $Wtb$ vertex
\begin{eqnarray}
\mathcal{L}_{Wtb} & = & - \frac{g}{\sqrt 2} \bar b \, \gamma^{\mu} \left( \vl P_L + \vr P_R
\right) t\; W_\mu^- \nonumber \\
& & - \frac{g}{\sqrt 2} \bar b \, \frac{i \sigma^{\mu \nu} q_\nu}{M_W}
\left( \gl P_L + \gr P_R \right) t\; W_\mu^- + \mathrm{h.c.} \,,
\label{ec:lagr}
\end{eqnarray}
where $\vl = V_{tb} + C_{\phi q}^{(3,3+3)*} \frac{v^2}{\Lambda^2}$, $\vr = \frac{1}{2} C_{\phi \phi}^{33*} \frac{v^2}{\Lambda^2}$, $\gl = \sqrt 2 C_{dW}^{33*} \frac{v^2}{\Lambda^2}$, $\gr = \sqrt 2 C_{uW}^{33} \frac{v^2}{\Lambda^2}$, being the gauge-invariant operators
\begin{align}
& O_{\phi q}^{(3,3+3)} = \frac{i}{2} \, \left[ \phi^\dagger (\tau^I D_\mu
  - \overleftarrow D_\mu \tau^I)  \phi \right] (\bar q_{L3} \gm \tau^I q_{L3}) \,,
&& {O_{\phi \phi}^{33}} = i (\tilde \phi^\dagger D_\mu \phi)
        (\bar t_{R} \gm b_{R}) \,, \notag \\
& O_{dW}^{33} = (\bar q_{L3} \smn \tau^I b_{R}) \phi \, \Wmn^I \,,
&& O_{uW}^{33} = (\bar q_{L3} \smn \tau^I t_{R}) \tilde \phi \, \Wmn^I \,,
\end{align}
in standard notation.
Were it not for this simplification, the $Wtb$ vertex would contain 6 additional Lorentz structures, with 6 more parameters arising from 5 (redundant) operators. Additional details can be found in Refs.~\cite{AguilarSaavedra:2008zc}.

One may always ask about precision electroweak data and constraints on these operators because, being the top quark the $\text{SU}(2)_L$ partner of the $b$ quark, gauge invariant operators containing the $t_L$ field automatically involve $b_L$. For example, the $Z b_L b_L$ vertex, well measured at LEP, receives corrections from effective operators
\begin{equation}
\mathcal{L}_{Z b_L b_L} = -\frac{g}{2 c_W} \bar b_L \gm \left[ -1 + \left( C_{\phi q}^{(3,3+3)}+C_{\phi q}^{(1,3+3)} \right) \frac{v^2}{\Lambda^2} +  \frac{2}{3} s_W^2 \right] b_L Z_\mu \,,
\end{equation}
where $O_{\phi q}^{(3,3+3)}$ is the operator producing a deviation $\delta V_L \equiv \vl - V_{tb}$ in the $Wtb$ vertex, and
$O_{\phi q}^{(1,3+3)} = \frac{i}{2} \, (\phi^\dagger \DMD \phi) (\bar q_{L3} \gm q_{L3})$. Does this imply that $C_{\phi q}^{(3,3+3)}$ is small, because fine-tuning with 
$C_{\phi q}^{(1,3+3)}$ (so as to keep agreement with LEP data) is unnatural? The answer is no. For example, mixing with a heavy charge $2/3$ isosinglet $T$, when integrated out,
precisely gives $C_{\phi q}^{(3,3+3)} = - C_{\phi q}^{(1,3+3)}$~\cite{delAguila:2000aa}.
And it is clear that it \emph{must} be so, because the mixing of $T$ with the SM charge $2/3$ mass eigenstates does not affect the down sector, where the GIM mechanism is still at work. When written in the language of effective operators, this is just the relation $C_{\phi q}^{(3,3+3)} = - C_{\phi q}^{(1,3+3)}$. Detailed analyses of the effects on $B$ physics of $Wtb$ anomalous couplings imply that $\vr$ and $\gl$ (that is, $C_{\phi \phi}^{33}$ and $C_{dW}^{33}$) are very constrained if no other new physics is present~\cite{Grzadkowski:2008mf}. But discarding operators merely because they contain terms with two $b$ fields is, as we can see, too naive and restrictive.

\section{The leading order approximation}

The anomalous couplings $\delta V_L $, $\vr$, $\gl$, $\gr$ in Eq.~(\ref{ec:lagr}) are proportional to the scale ratio $v^2/\Lambda^2$, and thus some expansion in powers of $1/\Lambda^2$ seems adequate. In general, the interference terms between SM amplitudes and dimension-6 operators are proportional to $1/\Lambda^2$, and the pure dimension-6 operator corrections to $1/\Lambda^4$. One might then naively consider keeping only the linear $1/\Lambda^2$ terms, which would give the ``leading'' corrections, and drop the quadratic $1/\Lambda^4$ ones. But this ``$1/\Lambda^2$'' approximation is far too restrictive, and does not allow to explore many new physics effects. Indeed, operators which do not interfere with the SM are the ones producing genuine new physics effects, beyond corrections to SM processes. One well-known example is given by top flavour-changing neutral processes, extremely suppressed in the SM, which are of order $1/\Lambda^4$. These processes, absent in the SM, could be visible for this reason even if suppressed by $1/\Lambda^4$.

\begin{figure}[t]
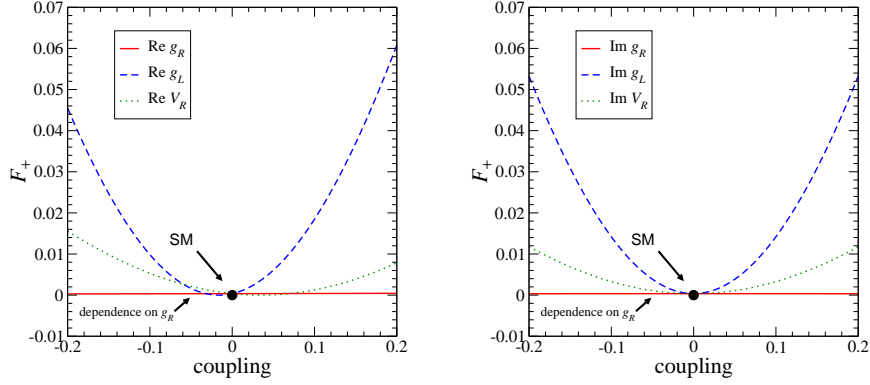

\begin{center}
\begin{tabular}{ccc}
\includegraphics[width=.35\textwidth]{Figs/FR-color.eps} & \quad &
\includegraphics[width=.35\textwidth]{Figs/FRi-color.eps}
\end{tabular}
\caption{Dependence of the helicity fraction $F_+$ on $Wtb$ anomalous couplings, for $\vl = 1$.}
\label{fig:FR}
\end{center}
\end{figure}

For the $Wtb$ vertex, in the $m_b=0$ limit (we will consider a non-zero $b$ mass later) only the operators generating $\bar b_L t_L$ and $\bar b_L t_R$ terms interfere with the SM amplitude. Then, the $1/\Lambda^2$ approximation leads us to keep only the corrections proportional to $\delta \vl$ and $\gr$ (see Table~\ref{tab:lagr}). This is unfortunate, because we lose the possibility of exploring $b$ chirality-breaking effects.%
\begin{table}[htb]
\begin{center}
\begin{tabular}{cccc}
& $1/\Lambda^2$ & \quad & $1/\Lambda^4$ \\
$\bar b_L \gm t_L$  & $\delta \vl$ && $(\delta \vl)^2+\;\text{dim-8}~ \bar b_L t_L$ \\
$\bar b_R \gm t_R$  & --                 && $(\vr)^2$ \\
$\bar b_R \smn t_L$ & --                 && $(\gl)^2$ \\
$\bar b_L \smn t_R$ & $\gr$        && $(\gr)^2+\;\text{dim-8}~ \bar b_L t_R$
\end{tabular}
\caption{Corrections to observables from $Wtb$ anomalous couplings, for $m_b = 0$.}
\label{tab:lagr}
\end{center}
\end{table}
It is well known that in SM top decays the fraction $F_+$ of $W$ bosons produced with helicity $+1$ is zero in the $m_b=0$ limit. The reason is that for the SM $Wtb$ vertex only left-handed $b$ quarks are produced in the decay, and then angular momentum conservation forbids $W$ bosons with helicity $+1$. (Including a $b$ quark mass $m_b = 4.8$ GeV the fraction of positive helicity $W$ bosons is $F_+ = 0.00359$.) However, the anomalous couplings $\delta \vl$, $\gr$ still involve left-handed $b$ quarks (otherwise they would not interfere with the SM amplitude) and their presence does not affect $F_+$. This can be seen in Fig.~\ref{fig:FR}, taken from Ref.~\cite{AguilarSaavedra:2010nx}. The (quadratic) corrections from anomalous couplings $\vr$, $\gl$ lead to dramatic departures from the SM prediction while a non-zero $\gr$ makes no difference. The sensitivity to these deviations in top decays at LHC is excellent~\cite{AguilarSaavedra:2006fy} precisely because $F_+$ is tiny within the SM, and in the $1/\Lambda^2$ approximation such effects are lost.

Let us then discuss how the $1/\Lambda^2$ approximation may be refined. As we see from Table~\ref{tab:lagr}, for $\bar b_L t_L$ and $\bar b_L t_R$ terms the linear corrections are of order $1/\Lambda^2$ while quadratic ones, as well as interference with dimension-8 operators with $\bar b_L t_L$, $\bar b_L t_R$ fields, are of order $1/\Lambda^4$. For $\bar b_R t_R$, $\bar b_R t_L$ fields the leading corrections are the quadratic $1/\Lambda^4$ ones $(\vr)^2$, $(\gl)^2$, while dimension-8 operators would give non-interfering corrections of order $1/\Lambda^8$. It is then consistent to keep for each $t,b$ chirality the lowest-order terms. (Sub-leading $(\delta \vl)^2$, $(\gr)^2$ terms cause no harm and may be kept or discarded at will.) This is what we denote as the ``leading order approximation''. When the $b$ quark mass is included, interferences of $\bar b_R t_R$ and $\bar b_R t_L$ operators are suppressed by $(m_b/m_t) 1/\Lambda^2$, and their effect is of the same order as the quadratic terms~\cite{AguilarSaavedra:2006fy}. Complete calculations of several top decay observables, keeping quadratic terms as well as the $b$ quark mass, have been performed in Refs.~\cite{AguilarSaavedra:2010nx,AguilarSaavedra:2006fy}, and for single top cross sections in Ref.~\cite{AguilarSaavedra:2008gt}.

\end{document}